\documentclass[pre,twocolumn]{revtex4}
\usepackage{amsmath,amscd,amsfonts,amssymb,color,float}
\usepackage{graphicx,amsfonts,epsf,hyperref}

\begin{document}


\title{QUANTUM BRAYTON ENGINE OF NON-INTERACTING FERMIONS IN   ONE DIMENSIONAL BOX
 } 



\author{Satnam Singh}
\email{satnamphysics@gmail.com, \\satnamsingh@iisermohali.ac.in}
\affiliation{Department of Physics, Rayat Bahra University
V.P.O. Sahauran, Tehsil Kharar, Distt. Mohali
Punjab, INDIA - 140104}


\date{\today}

\begin{abstract}
We consider the quantum Brayton cycle without temperature, constructed from non-interacting fermions, trapped in a one-dimensional box.  We use all energy levels of the box. The work and heat in this cycle are calculated from the expectation values of the Hamiltonian. We analytically calculated the efficiency of the cycle, efficiency at maximum work and Clausius relation as the function of the ratio of the lengths. 
We found that the efficiency of the cycle does not depend on the number of fermions. It depends on the ratios of the lengths of the cycle, while the power depends on the number of fermions. The irreversibility of the cycle also does not depend on the number of particles. It only depends on the ratio of the box lengths. Moreover, We also analysed the relation of efficiency and the power of the cycle. We found that as we decrease the ratio of the lengths, the efficiency at maximum power and the  maximum power of the cycle increases. The power of the cycle  increases as we increase the number of the particles.

\end{abstract}


\maketitle 

\section{INTRODUCTION}
Heat engine is a well-known example of thermal machine\cite{spin-quantum-SA,zhang2014quantum,seah2018work,alicki2015non,correa2014optimal,ramezani2019impact,abe2011maximum,singh2018low}.
It converts the heat energy into mechanical work\cite{mani2018graphene,huang2012effects}.
These thermal machines undoubtedly play an essential role in our everyday life.
 Cars, refrigerators, air-conditioners, lasers and power plants are the most common examples of thermal machines, which are directly or indirectly involved in daily life.
In the modern era, technologies being able to miniaturization things down to the nanoscale where quantum effects may not be
negligible. 
So it is essential to study these quantum systems directly in relation to  thermodynamics system\cite{dattagupta2017ericsson,rossnagel2014nanoscale,uzdin2015equivalence,uzdin2016quantum,abe2013general,chen2019achieve,singh2019three}.

In the early 19th century, 
Sadi Carnot gave a mathematical model of reversible and cyclic heat engine. That model consisted of a  cylinder with a piston filled with the ideal gas.  The efficiency of that was independent of the working substance used in the engine. It only depends on the temperature of the external cold($T_c$) and hot temperature$T_h$ reservoirs. Efficiency of that cycle is written as $\eta_c=1-T_c/T_h$ \cite{bender2002entropy,bender2000quantum}.
In the modern age, technologies can miniaturize things down
to the nanoscale level where quantum effects are not
negligible\cite{dattagupta2017ericsson,rossnagel2014nanoscale}.
These effects influence the thermodynamical properties of the system dramatically.
In this regime, the traditional classical thermodynamics based on macroscopic bodies does not remain applicable. 
It created more interest in the applicability of thermodynamics in the quantum regime.
Various studies are available which describe the quantum version of thermodynamics. In these studies, all processes of the cycle formulated corresponding to their classical analogue\cite{deffner2010quantum,deffner2016quantum}.

In quantum heat engines, the quantum
matter is used as the working substance\cite{yin2017optimal,bender2000quantum,chand2017single,deffner2018efficiency,rossnagel2014nanoscale,jiang2018near,humphrey2002reversible,chen2018bose,thomas2019quantum,singh2017feynman}. The fundamental conceptual difference between these classical and quantum thermal machines is that in the quantum thermal machines one is concerned
with the discrete energy levels of particles. In the
literature, various examples of quantum thermal machines can
be found\cite{watanabe2017quantum,agarwalla2017quantum,kieu2004second,goswami2013thermodynamics,singh2018feynman,thomas2012informative}. These examples include the quantum analogues of Brayton, Diesel, Lenoir, Carnot, Stirling and Otto cycles\cite{wang2013efficiency,latifah2013quantum,sutantyo2015quantum,single-ion-obina,harbola2012quantum}.
In these cycles, different quantum systems, such as non-interacting harmonic oscillator systems, particle in a box, spin systems, and two-level or multilevel systems are used as working substance\cite{quan2009quantum,quan2007quantum,rossnagel2016single,geva1992quantum,wang2018nonlinear}.
These quantum heat engines has  
received more attention than the classical one,  because they  are inherently clean and thermally more efficient than the classical heat engines.
The  Scovil and Schultz-Dubois are the pioneers who introduced the concept of the quantum heat engine for the
first time. They used three level maser as a quantum heat engine\cite{scovil1959three,dorfman2018efficiency}.
After that a lot of papers has published in this field\cite{thomas2017implications,lu2019optimal,humphrey2005quantum,zhang2014quantum,gardas2015thermodynamic}.

 In 2000 Bender et al. gave a mathematical model of infinite time Carnot cycle, whose working substance was a particle trapped in a square well potential\cite{bender2002entropy,bender2000quantum}. That cycle consists of four steps, adiabatic expansion, isoenergetic expansion, adiabatic compression and isoenergetic compression. These steps were analogues to the classical  Carnot cycle. After this lot of papers related to the particle in a box heat engine have been published\cite{latifah2013quantum,wang2012optimization,latifah2014multiple}.
 
 Brayton heat engine was originally proposed by John Barber in 1791, but it was named after  George Brayton. 
Here, we are going to present Brayton heat engine, whose working substance is fermi gas, trapped in one-dimensional box. In this paper, we calculate the efficiency of the cycle, efficiency at maximum work and Clausius relation of the cycle. We studied the nature of these quantities as the function of the ratio of the lengths.
We also analysed relationship between the efficiency and the power of the cycle.

\section{QUANTUM THERMODYNAMICS OF PARTICLE IN BOX}
Consider that a single  particle trapped in a one-dimensional box.
 The  energy  the $i$th level of the system is written as\cite{wang2012performance,latifah2013quantum}
\begin{equation}
 \epsilon_i=\frac{\pi^2\hbar^2 i^2}{2mL^2} \label{eq:energy-i}
\end{equation}

Let $p_i$ is the probability of finding the particle in $i$th  level. The average energy of the system is written as

\begin{equation}
U=\sum_i \frac{\pi^2\hbar^2 i^2}{2mL^2} p_i= \sum_i p_i \epsilon_i \label{eq:internal-energy}
\end{equation}

 Equation (\ref{eq:internal-energy}) corresponds to the internal energy of the system.
 When the system  is at  thermal contact with the  reservior, the internal energy of the system changes and the system performs work. The first law of the thermodynamics is written as
 \begin{equation}
  dU=dQ-dW
 \end{equation}
Differentiating equation (\ref{eq:internal-energy}) we have\cite{schrodinger1989statistical}
\begin{equation}
 dU=\sum_i(\epsilon_i dp_i+ p_i d \epsilon_i )
\end{equation}
Due to the relationship $S=-k_B \sum p_i ln p_i$ between the entropy($S$) and the probability($p_i$) the above equations become
\begin{equation}
 dQ=\sum_i \epsilon_i d p_i, \hspace{0.25cm} \hspace{0.25cm} -dW=\sum_i p_i d\epsilon_i \label{eq:basic-eq}
\end{equation}
The above equations  are corresponding to heat and work done in the particle in box heat engines, which are analogues to classical definition of heat and work.

  \begin{figure}
 \centering
 \includegraphics[scale=0.65,keepaspectratio=true]{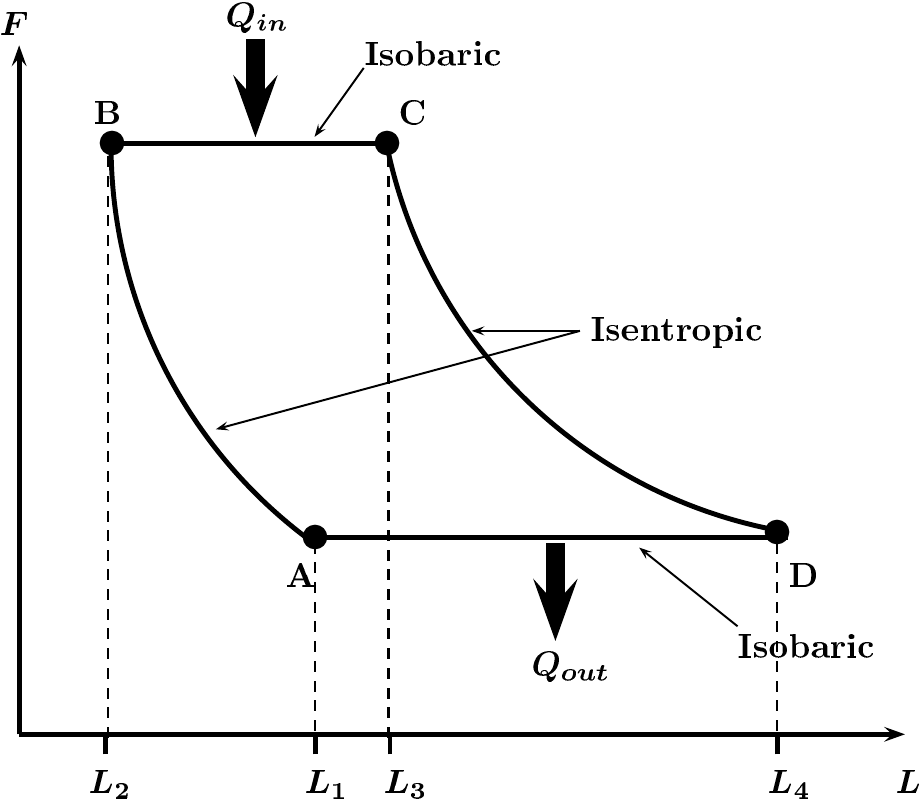}
 \caption{This figure represents the Brayton cycle. This cycle consists of four steps v.i.z adiabatic compression($A$ to $B$), isobaric expansion ($B$ to $C$), adiabatic expansion($C$ to $D$) and isobaric compression($D$ to $A$)} \label{fig:cycle}
\end{figure}

\section{FERMION QUANTUM HEAT ENGINE}
Brayton cycle is a four-step cycle. It consists of two isentropic and two isobaric steps. These steps are given in figure \ref{fig:cycle}.
In this work, we consider one-dimensional box having length $L$ and filled with $N$ number of fermions of mass $m$. Due to Pauli's exclusion principle, two fermions can not occupy the same quantum state of the box. Initially we consider that the first fermion is in $j_{1}$ state, second fermion is in $j_2$ state ...so on, and the $N^{th}$ fermion is in $j_N$ state. Generally the $i$th fermion is in $j_i$ state. All $j$ are different and have following order $j_1<j_2<j_3<...j_N.$  
  The energy eigen value($\epsilon_{j_i}$) of the $j_i^{th}$ energy level is given as
  \begin{equation}
  \epsilon_{j_i}=\frac{\pi^2\hbar^2 }{2mL^2} j_i^2  
 \end{equation}
  
   The  total energy of the system is written as
 \begin{equation}
  U=\sum_{i=1}^N \epsilon_{j_i}(L)=\frac{\pi^2\hbar^2 }{2mL^2}\sum_{i=1}^N j_i^2  
 \end{equation}
 We assume that one of the infinite wall of the potential is allowed to move infinitesimal distance. Then the  eigen function and the energy eigen value vary as the function of $L$. 
 The force during each thermodynamical process is calculated as $F=-\partial E/\partial L$ and given by
\begin{equation}
 F=\frac{\pi^2\hbar^2 }{mL^3}\sum_{i=1}^N j_i^2 
\end{equation}

\begin{figure}
 \centering
 \includegraphics[scale=0.65,keepaspectratio=true]{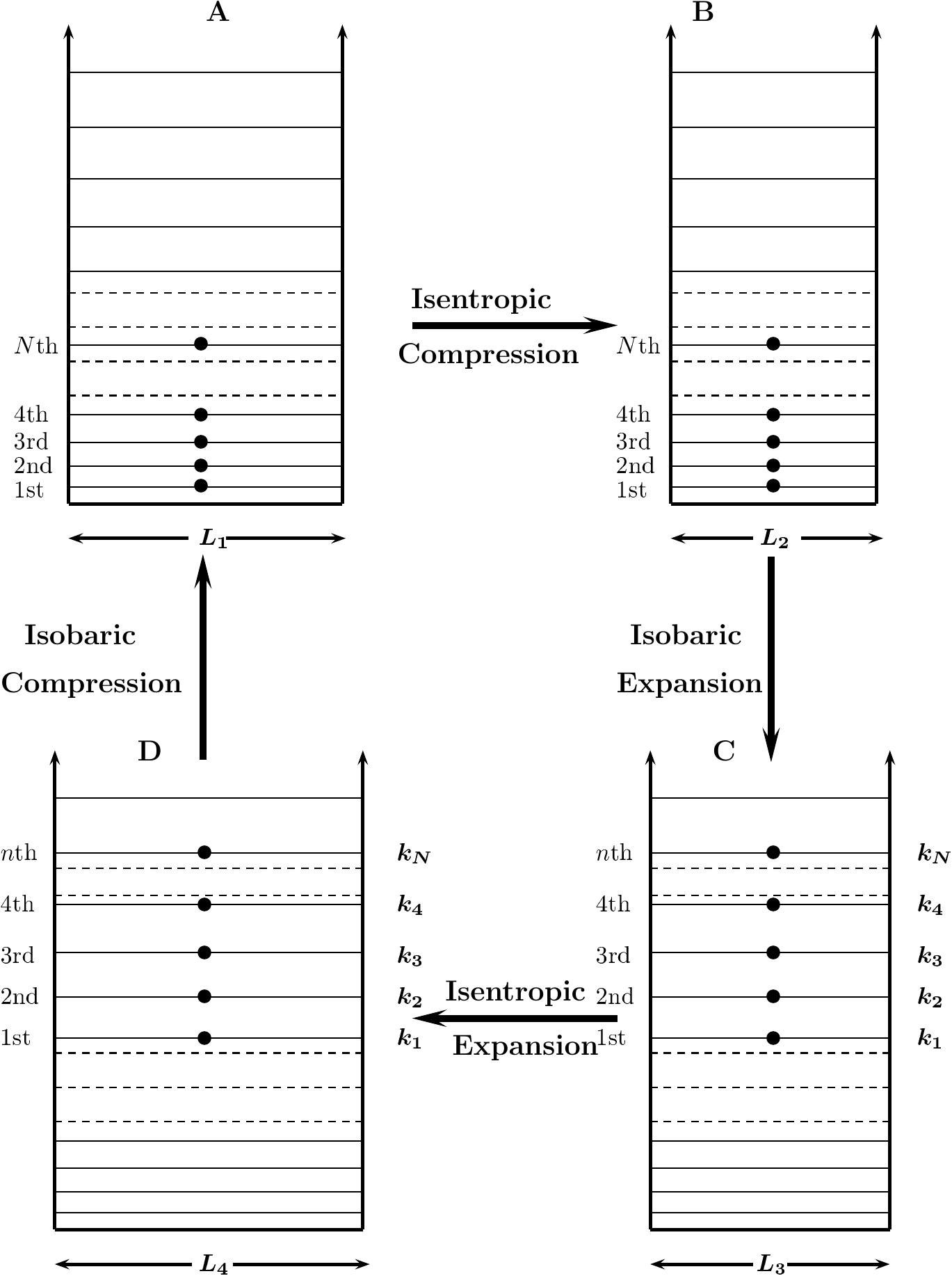}
 \caption{This figure shows the pictorial representation of particle in box Brayton cycle} 
\end{figure}

Now we will discuss the all steps performed in the cycle.

\subsubsection*{ISENTROPIC COPMRESSION}

In isentropic compression ($A$ to $B$), particles remain in the same state. The length of the box decreases. There is no heat exchange during this step. The first law of thermodynamics is written as
 \begin{equation}
  dU=-dW
 \end{equation}
The change in internal energy is used to perform work. 
The box length is reduced to $L_2$ from $L_1$. The work done on the system is written as
\begin{equation}
 W_{AB}=-\frac{\pi^2 \hbar^2 }{2mL_2^2}\Big(1-\frac{L_2^2}{L_1^2}\Big)\sum_{i=1}^N j_i^2
\end{equation}
We assign $S_i=\sum_{i=1}^N j_i^2$. 

\subsubsection*{ISOBARIC  EXPANSION}
In isobaric expansion($B$ to $C$), particles jumps to the higher states.  The heat is added in this step. Due to the constant pressure, the whole process performed with constant force($F_B=F_C$).  The first law of thermodynamics is written as
 \begin{equation}
  dU=dQ_{in}-F_{cons} \int_B^C dL
 \end{equation}
Where $F_{cons}$ is the constant force.
 Length of the system is expanded to $L_3$ from $L_2$. Particles jump to the higher energy states. Initially, the first particle was $j_1$ state, now it jumps to the $k_1$ state, second particle to $k_2$ state and $N^{th}$ particle to $k_N$ state, where all $k$s are different. 
 Generally the fermion in the $j_i$ state jump to the $k_i$ state. Where  $k_i>j_i$.
 The force at point $B$ and at point $C$ are written as
 \begin{equation}
  F_{B}=\frac{\pi^2\hbar^2 }{mL_2^3}\sum_i^N j_i^2, \hspace{1cm}   F_{C}=\frac{\pi^2\hbar^2 }{mL_3^3}\sum_i^N k_i^2
 \end{equation}

 By equating the forces($F_B=F_C$) \cite{wang2012performance,latifah2013quantum}, we get
\begin{equation}
 \frac{\sum_{i=1}^N j_i^2}{L_2^3}=\frac{\sum_{i=1}^N k_i^2}{L_3^3}\label{eq:ratio-1}
\end{equation}
We assign $S_f={\sum_{i=1}^N k_i^2}$.
 The work done by the system is written as

\begin{equation}
 W_{BC}=\frac{\pi^2 \hbar^2 S_i}{2mL_2^2}\Big(\frac{2 L_3}{L_2}-2\Big)
\end{equation}
The change in the internal energy is written as

\begin{equation}
 U_{BC}=\frac{\pi^2 \hbar^2 }{2m}\Big(\frac{S_f}{L_3^2}-\frac{S_i}{L_2^2}\Big)
\end{equation}

The heat added in the system is calculated from the first law of thermodynamics, which is written  as $Q_{in}=U_{BC}+W_{BC}$. The heat added in the system is written as 
 \begin{equation}
  Q_{in}=-\frac{3\pi^2 S_i(\alpha-1) \hbar^2}{2m L_2^2 \alpha} \label{eq:qin}
 \end{equation} 
 Where $\alpha$ is the ratio of $L_2$ and $L_3$.
 
 \subsubsection*{ISENTROPIC EXPANSION}
 During isentropic expansion($C$ to $D$), the length of the box is expanded to $L_4$ from $L_3$ and particles do not change their quantum states. There is no heat exchange during this step. 
  The work done by the system is written as

\begin{equation}
 W_{CD}=\frac{\pi^2 \hbar^2 S_f}{2m}\Big(\frac{ 1}{ L_3^2}-\frac{ 1}{L_4^2 }\Big),
\end{equation}
where $ W_{CD}$ is the work done during isentropic expansion.
\subsubsection*{ISOBARIC COPMRESSION}
The system is compressed in isobaric manner from $D$ to $A$. In this step, particles return to their initial states  and the heat is dumped out ($Q_{out}$) from the system.  Force  remains constant during this process($F_D=F_A$). Whole system returns to its initial condition. The force at point $D$ and at point $A$ are written as
 \begin{equation}
  F_{D}=\frac{\pi^2\hbar^2 }{mL_4^3}\sum_i^N k_i^2, \hspace{1cm}   F_{A}=\frac{\pi^2\hbar^2 }{mL_1^3}\sum_i^N j_i^2
 \end{equation}

Equating both forces, we get
\begin{equation}
 \frac{\sum_{i=1}^N j_i^2}{L_1^3}=\frac{\sum_{i=1}^N k_i^2}{L_4^3} \label{eq:ratio-2}
\end{equation}
The workdone on the system is written as 
\begin{equation}
 W_{DA}=\frac{\pi^2 \hbar^2 S_i}{mL_1^3}\Big(L_1-L_4)
\end{equation}
Total change in internal energy of the system is written as
\begin{equation}
 U_{DA}=\frac{\pi^2 \hbar^2 }{2m}\Big(\frac{S_i}{L_1^2}-\frac{S_f}{L_4^2}\Big)
\end{equation}
From equation (\ref{eq:ratio-1}) and (\ref{eq:ratio-2}), we can write
\begin{equation}
 \frac{S_i}{S_f}=\Big(\frac{L_2}{L_3}\Big)^3=\Big(\frac{L_1}{L_4}\Big)^3=\alpha^3
\end{equation}

 Heat dumped out by the system is calculated as $Q_{out}=U_{DA}+W_{DA}$, and is given as
\begin{equation}
 Q_{out}=\frac{3\pi^2 S_i(\alpha-1) \hbar^2}{2m L_4^2 \alpha^3} \label{eq:qout}
\end{equation}

 Total work done by the system is calculated from the sum of all work done in the cycle  $W=W_{AB}+W_{BC}+W_{CD}+W_{DA}$, which is written as
 \begin{equation}
  W=\frac{3 \pi^2 S_i \hbar^2(\alpha-1)(L_2-L_4 \alpha)(L_2+L_4 \alpha)}{2m L_2^2 L_4^2 \alpha^3} \label{eq:work}
 \end{equation}
 
 The efficiency of the Brayton cycle($\eta$) is the ratio of total work done by the system and the heat input. The efficiency is written as
\begin{equation}
 \eta=\frac{W}{Q_{in}}=1-\frac{L_2^2}{L_4^2 \alpha^2} \label{eq:eta}
\end{equation}

The equation (\ref{eq:eta}) refers to the efficiency of the Brayton cycle as the function of the box lengths. It is clear that it does not depend on the number of fermions in the system.

\section{EFFICIENCY AT MAXIMUM WORK}
Now we are able to calculate the efficiency at maximum work of the cycle. The work of the cycle is given in equation \ref{eq:work}. The work done by one cycle is a function of all lengths of the engine. Where
 $L_2$ is the minimum length of the box and $L_4$ is the maximum length of the box. For a particular engine, minimum and maximum limits of  lenght are fixed. So in this work, both  lenghts are considered constant.
Differentiating $W$ with $\alpha$ and equating it with zero. The value of the  $\alpha$  is written as

\begin{equation}
 \alpha=\frac{-L_2^2+\sqrt{ L_2^4 +3L_4^2 L_2^2}}{ L_4^2}
\end{equation}
Substituting the value of $\alpha$ in equation (\ref{eq:eta}), we will get efficiency at maximum work($\eta_{mw}$), which is given as
\begin{equation}
 \eta_{mw}=1-\frac{L_2^2 L_4^2 }{(-L_2 ^2+\sqrt{L_2^4+3 L_2^2 L_4^2})^2} \label{eq:emp}
\end{equation}

We consider that $L_2/L_4=\beta$. The value of the $\beta$ lies between 0 and 1.
\begin{equation}
 \eta_{mw}=\frac{2}{9} (3-\beta^2-\beta \sqrt{3+\beta^2})\label{eq:efmp}
\end{equation}

\begin{figure}
 \centering
 \includegraphics[scale=0.65,keepaspectratio=true]{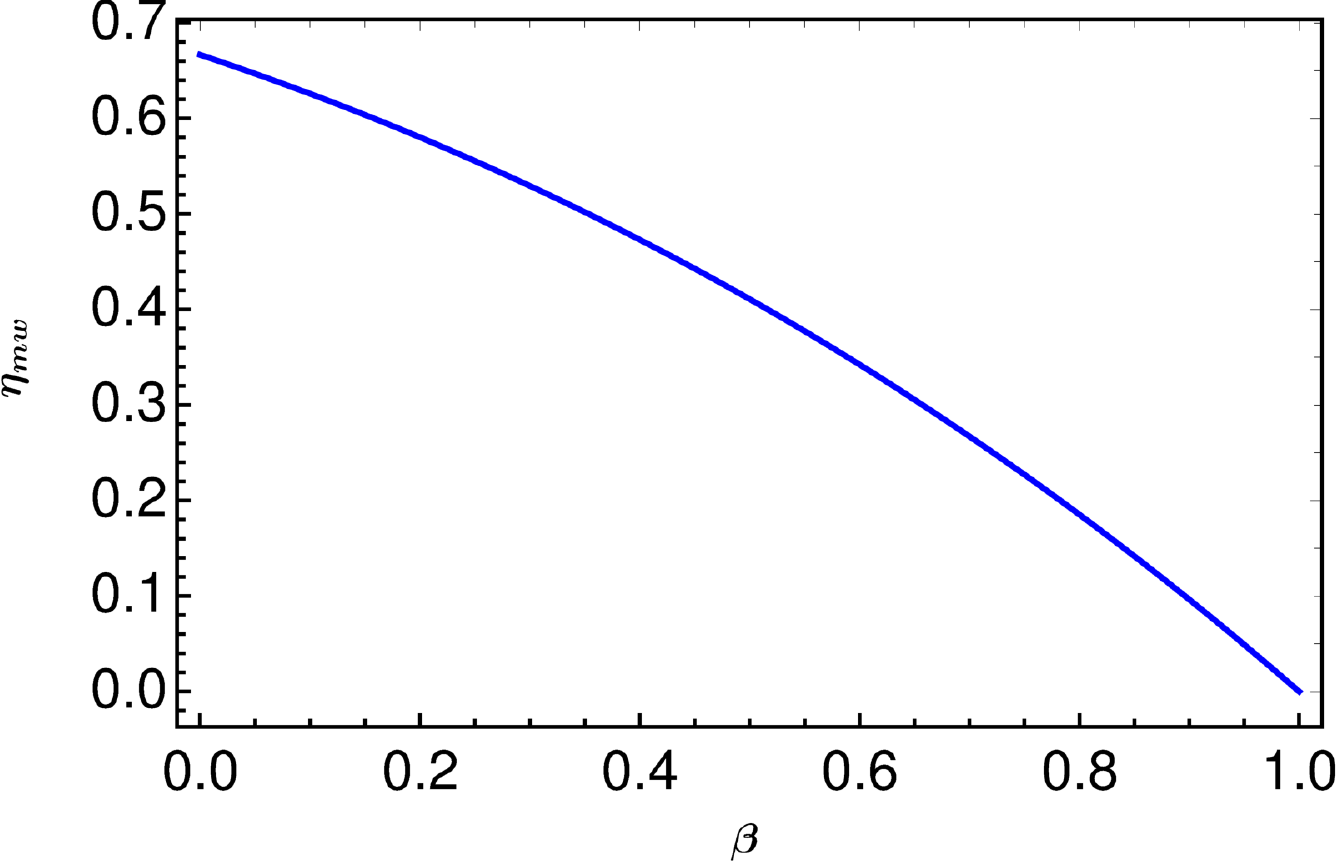}
 \caption{Efficiency at maximum power as the ratio of $L_2$ and $L_4$} \label{fig:emp}
\end{figure}

The equation (\ref{eq:efmp}) represents the efficiency at maximum work as the function of the ratio of $L_2$ and $L_4$. It is plotted in the figure (\ref{fig:emp}). It suggests that as we increase the ratio of the lengths($\beta$), the efficiency at maximum work will decrease.

\section{CLAUSIUS RELATION FOR PARTICLE IN
BOX BRAYTON ENGINE}

Rudolf Clausius derived a mathematical relation to distinguishing between reversible and irreversible heat cycles. That relation is known as Clausius relation\cite{leff2018reversible}. It is written as
\begin{equation}
 \oint \frac{dQ}{T}\leq0
\end{equation}
Where $Q$ is heat, and $T$ is temperature. In the case of the reversible cycle, it holds equality, but in the case of the irreversible cycle, it holds the inequality. For the particle in box heat engines, Bender et al. proposed a relation analogues to the Clausius relation\cite{bender2002entropy}, which is written as
\begin{equation}
 \oint \frac{dQ}{E}\leq0
\end{equation}
Although this relation can not be used to determine the entropy change of the cycle, but it gives the insight about the reversibility of the cycle.
This heat engine operated between two extreme energy limits, which are given as
\begin{equation}
 E_H=\frac{\pi^2 \hbar^2 S_f}{2 m L_3^2}, \hspace{0.5cm}  \hspace{0.5cm} E_C=\frac{\pi^2 \hbar^2 S_i}{2 m L_1^2}
\end{equation}
 \begin{figure}
 \centering
 \includegraphics[scale=0.65,keepaspectratio=true]{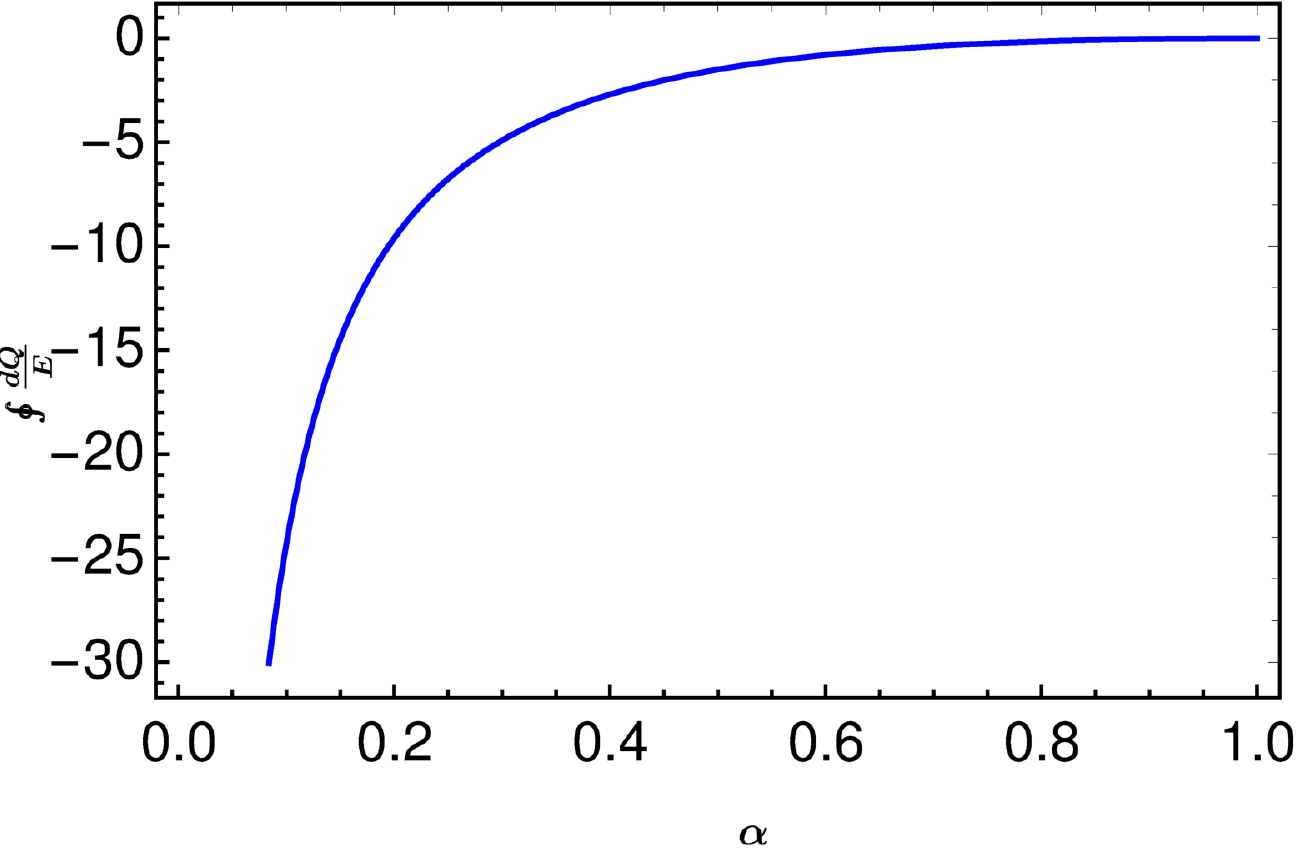}
 \caption{ This figure shows the relation between $\oint \frac{dQ}{E}$ and $\alpha$. It is clear that the irreversible of the cycle does not depend on the number of particles in the box. } \label{fig:ine}
\end{figure}
The input heat and the  dumped out heat are given in the equations (\ref{eq:qin}) and (\ref{eq:qout}) respectively.
The above relation is written as

\begin{equation}
 \oint \frac{dQ}{E}=\frac{Q_{in}}{E_H}+\frac{Q_{out}}{E_C}=  -\frac{3(-1+\alpha)^2}{\alpha} \label{eq:ineq}
\end{equation}
Equation (\ref{eq:ineq}) gives the relation between $\oint \frac{dQ}{E}\leq0$ and $\alpha$. It is plotted in figure (\ref{fig:ine}). This shows that the cycle is   irreversible. The irreversibility the  cycle depends only on the ratio of the lengths $L_2$ and $L_3$ or $L_1$ and $L_4$. It does not depend on the number of the particles. As  we increase the value of the $\alpha$, the irreversibility of the cycle decreases.

\section{OPTIMIZATION ON THE PERFORMANCE OF THE
HEAT ENGINE}
Now we are going to discuss about the power output and efficiency of the quantum Brayton cycle \cite{wang2012performance,wang2012optimization,abe2011maximum,abe2013general,abe2012role}.  Let the $v$ is the avarage speed of the variation of length of the box. The total change in the length in one cycle is witten as
\begin{equation}
 L=(L_1-L_2)+(L_3-L_2)+(L_4-L_3)+(L_4-L_1)=2(L_4-L_2)
\end{equation}

Where $L$ is the total change in length in one cycle. The power of the cycle is written as
\begin{equation}
 P=\frac{W}{\tau} \label{eq:pow}
\end{equation}

The avarage speed is written as
\begin{equation}
 v=\frac{L}{\tau}=2\frac{L_4-L_2}{\tau}
\end{equation}

To apply the adiabatic theorem, the average speed ($v̄$) must satisfy the 
condition that $v<<\frac{L}{\hbar/E}$\cite{wang2012performance,wang2012optimization}.
The power of the heat cycle is given in the equation \ref{eq:pow}. The total cycle time is $\tau= 2(L_4-L_2)/v$, then the power is written as

\begin{equation}
 P=\frac{3 \pi^2 S_i v(-1+\alpha)(L_2-L_4 \alpha)(L_2+L_4 \alpha)\hbar^2}{4 L_2^2 L_4^2 (-L_2+L_4) m \alpha^3} \label{eq:powcycle}
\end{equation}
Equation (\ref{eq:powcycle}) represents the power of the Brayton cycle as the function of lengths, $\alpha$ and the mass of the particle. From this equation, we can investigate the relation of the power and the number of the particles. It can be written as
\begin{equation}
 P\propto S_i \label{eq:pow-N}
\end{equation}
The equation \ref{eq:pow-N} shows that the power of the cycle increases as we increase the number of the particles.
Now we have to deduce the relation between efficiency and the power of the cycle. Using equation \ref{eq:eta}, the power of the cycle is written as
\begin{equation}
 P=\frac{3 \pi^2 S_i v (-1+\alpha) \sqrt{1-\eta} \eta \hbar^2 }{4 L_2^3 m (-1+\alpha \sqrt{1-\eta})} 
\end{equation}

\begin{figure}
 \centering
 \includegraphics[scale=0.65,keepaspectratio=true]{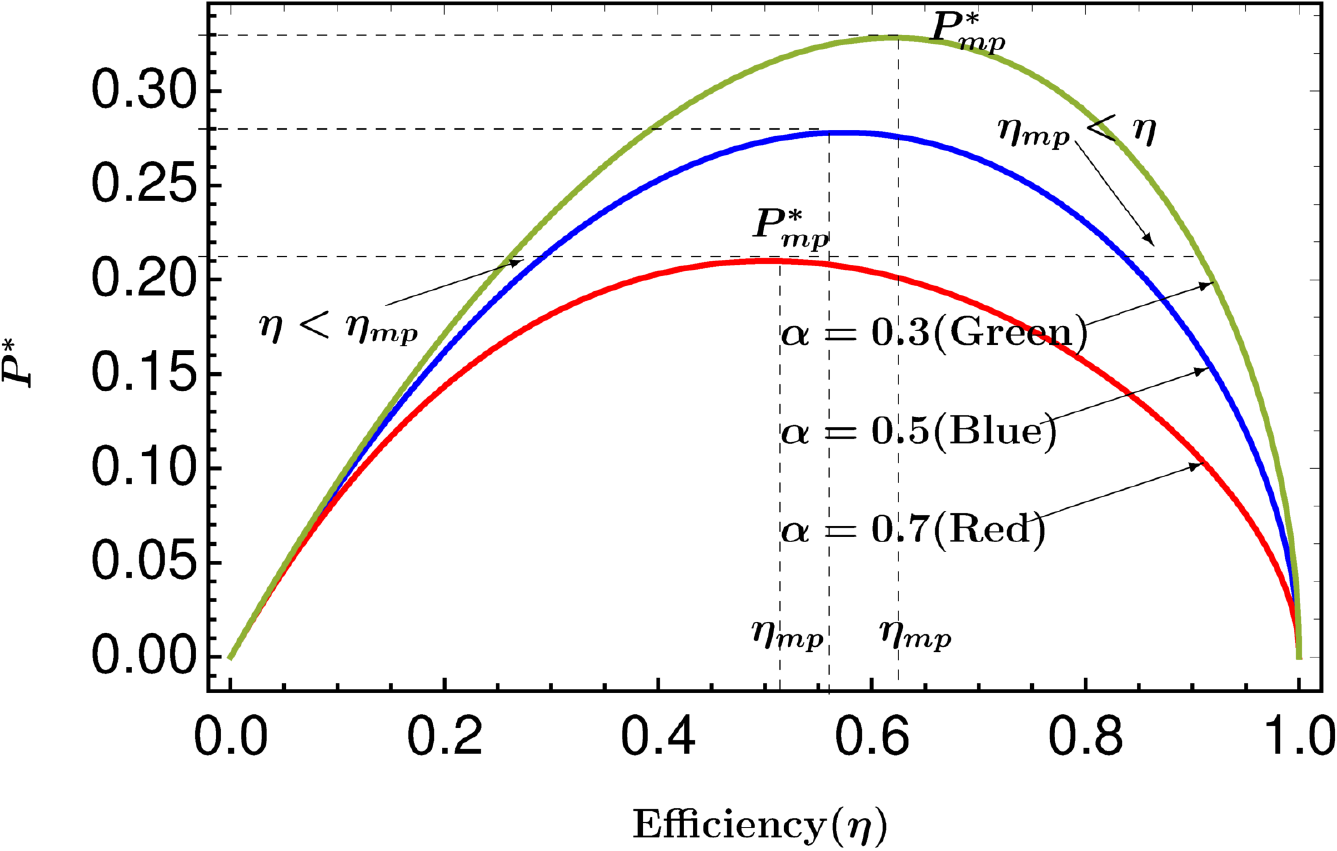}
 \caption{Plot of dimensionless power($P^*$) and efficiency of the cycle at different values of $\alpha$. Where Green curve is for $\alpha=0.3$, blue curve for $\alpha=0.5$ and red curve for $\alpha=0.7$ respectively.} \label{fig:dmlspow}
\end{figure}

Now we are able to calculate the dimensionless power as the function of the ratio of the lengths and the efficiency of the cycle.
We assign $K=3 \pi^2 S_i v \hbar/4L_2^3 m$. The dimensionless power($P^*$) is written as
\begin{equation}
 P^*=\frac{P}{K} =\frac{(-1+\alpha) \sqrt{1-\eta} \eta  }{ (-1+\alpha \sqrt{1-\eta})} \label{eq:dmlspow}
\end{equation}
Equation (\ref{eq:dmlspow}) represents the relation between dimensionless power, efficiency and the ratio of the lengths of the engine. For the different values of the $\alpha$, dimensionless power and efficiency is plotted in the figure (\ref{fig:dmlspow}).

The figure (\ref{fig:dmlspow}) shows that the characteristic curve of $P^*-\eta$ is parabola like. As we increase the value of the $\alpha$, the maximum power output decreases.  Corresponding to that maximum power, the efficiency($\eta_{mp}$) also decreases. 

 \begin{figure}
 \centering
 \includegraphics[scale=0.65,keepaspectratio=true]{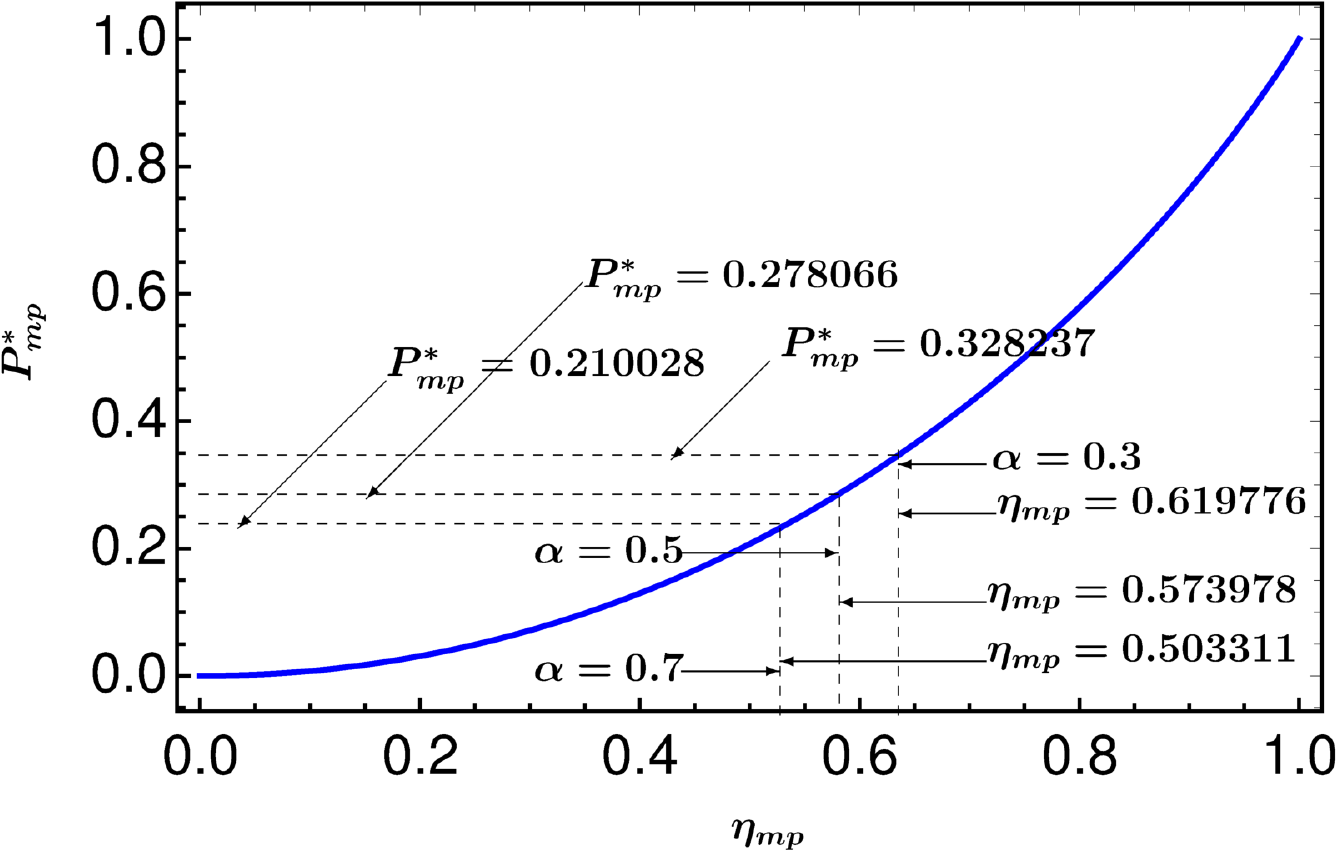}
 \caption{Plot of maximum dimensionless power ($P^*_{mp}$) and efficiency at maximum power ($\eta_{mp}$). As we increase the value of the $\alpha$, the efficiency at maximum power and the maximum power of the cycle decreases.} \label{fig:eff-dmlspow}
\end{figure}

\begin{table}
\caption{In this table efficiency at maximum power and maximum dimensionless power for different values of $\alpha$ is given.}
\renewcommand{\arraystretch}{0.95}
\centering
\begin{tabular}{|c|c|c|c|c|}
\hline
  $\alpha$ &  $\eta_{mp}$&$P^*_{mp}$\\ \hline 
  0.3&0.619776&0.328237\\
  0.5&0.573978&0.278066\\
  0.7&0.503311&0.210028\\
     \hline
\end{tabular}\label{tab:eff-at-max-pow}
\end{table}

For the different value of $\alpha$, there is a different set of values  of efficiency at maximum power and maximum power.
Now we will discuss about the relation between  maximum power and efficiency at maximum power for different values of $\alpha$.  At maximum dimensionless power,  the relation between $\alpha$ and $\eta$  can be calculated by equating the  first derivative of dimensionless power with $\alpha$ to zero, which is given as
\begin{equation}
 \alpha=\frac{2-3\eta}{2(1-\eta)^{3/2}}
\end{equation}
Substituting the value of $\alpha$ in equation (\ref{eq:dmlspow}), we get the relation between 
maximum dimensionless power and  efficiency at maximum power, which is given as

\begin{equation}
 P^*_{mp}=2 (1-\eta_{mp} )^{3/2}  +3 \eta_{mp} -2
\end{equation}
The figure (\ref{fig:eff-dmlspow}) shows the plot between maximum dimensionless power and  and efficiency at maximum power.In table (\ref{tab:eff-at-max-pow}),  values of the maximum dimensionless power and efficiency at maximum power at different values of the $\alpha$ is given.
From figure \ref{fig:eff-dmlspow} and table \ref{tab:eff-at-max-pow}, it is clear that as we decrease the value of $\alpha$, efficiency at maximum power and the maximum power output increases. We can enhanse the 
 power of the cycle  by increasing the value of $\alpha$.

\section{CONCLUSIONS}
We successfully studied the Brayton cycle constructed from the fermions in a one-dimensional box. We found that the efficiency of the cycle depends on the ratio of the lengths and it is independent from the number of particles. The efficiency of the maximum work also depends on the ratio of the lengths.  The irreversibility of the cycle,  also does not depend on the number of particles.
Characteristic of the dimensionless power and the efficiency is parabola like.
 As we increase the value of the $\alpha$, the efficiency at maximum power ($\eta_{mp}$) and corresponding maximum power($P^*_{mp}$) decreases. Power of the cycle can be increases by increase in the  $\alpha$
\section{ACKNOWLEDGEMENT}
SS gratefully acknowledge insightful discussions
with Obinna Abah.
\bibliography{brayton}
\bibliographystyle{apsrev4-1}

\end{document}